\pdfoutput=1
\documentclass[journal, 10pt]{IEEEtran}

\IEEEoverridecommandlockouts
\usepackage{amsmath,amssymb,amsfonts,bm}

\usepackage{algorithmic}
\usepackage{fancyhdr, graphicx}
\usepackage{tikz}
\usetikzlibrary{mindmap}
\usetikzlibrary{calc}
\usetikzlibrary{shapes}
\usepackage{textcomp}
\usepackage{xcolor, colortbl}
\usepackage[utf8]{inputenc}
\usepackage[english]{babel}
\usepackage[autostyle]{csquotes}
\usepackage[ruled,vlined]{algorithm2e}
\usepackage{csvsimple}
\usepackage{bbm}
\usepackage{caption}
\usepackage{balance}
\usepackage{comment}
\usepackage{subfiles}
\usepackage{cite}
\usepackage{paralist}
\usepackage{booktabs}
\usepackage{tabu}
 
\definecolor{Gray}{gray}{0.8}
\definecolor{Gray2}{gray}{0.9}

\usepackage{acro}

\DeclareAcronym{AMC}{
  short = AMC ,
  long  = Automatic Modulation Classification ,
  class = abbrev
}

\DeclareAcronym{DL}{
  short = DL ,
  long  = deep learning ,
  class = abbrev
}

\DeclareAcronym{V2V}{
  short = V2V ,
  long  = Vehicle-to-Vehicle ,
  class = abbrev
}

\DeclareAcronym{HMI}{
  short = HMI ,
  long  = Human-Machine Interaction ,
  class = abbrev
}

\DeclareAcronym{V2I}{
  short = V2I ,
  long  = Vehicle-to-Infrastructure ,
  class = abbrev
}

\DeclareAcronym{RNN}{
  short = RNN ,
  long = Recurrent Neural Network ,
  class = abbrev
}

\DeclareAcronym{IoT}{
  short = IoT ,
  long = Internet-of-Things ,
  class = abbrev
}

\DeclareAcronym{UAV}{
  short = UAV ,
  long = Unmanned Aerial Vehicle ,
  class = abbrev
}

\DeclareAcronym{SEI}{
    short = SEI ,
    long = Specific Emitter Identification ,
    class = abbrev
}

\DeclareAcronym{AE}{
    short = AE ,
    long = Autoencoder ,
    class = abbrev
}

\DeclareAcronym{CNN}{
    short = CNN ,
    long = Convolutional Neural Network ,
    class = abbrev
}

\DeclareAcronym{AI/ML}{
  short = AI/ML ,
  long  = Artificial Intelligence/Machine Learning ,
  class = abbrev
}

\DeclareAcronym{AI}{
  short = AI ,
  long  = Artificial Intelligence ,
  class = abbrev
}

\DeclareAcronym{NLP}{
  short = NLP ,
  long  = Natural Language Processing ,
  class = abbrev
}

\DeclareAcronym{RFML}{
  short = RFML ,
  long  = Radio Frequency Machine Learning ,
  class = abbrev
}

\DeclareAcronym{RC}{
  short = RC ,
  long  = Reservoir Computing ,
  class = abbrev
}

\DeclareAcronym{MIMO}{
  short = MIMO ,
  long  = multiple-input-multiple-output ,
  class = abbrev
}

\DeclareAcronym{SISO}{
  short = SISO ,
  long  = single-input-single-output ,
  class = abbrev
}

\DeclareAcronym{OFDM}{
  short = OFDM ,
  long  = orthogonal frequency-division multiplexing ,
  class = abbrev
}

\DeclareAcronym{DARPA}{
  short = DARPA ,
  long  = Defense Advanced Research Projects Agency ,
  class = abbrev
}

\DeclareAcronym{IQ}{
  short = IQ ,
  long  = In-Phase and Quadrature ,
  class = abbrev
}

\DeclareAcronym{DSA}{
  short = DSA ,
  long  = Dynamic Spectrum Access ,
  class = abbrev
}

\DeclareAcronym{GAN}{
  short = GAN ,
  long  = Generative Adversarial Network ,
  class = abbrev
}

\DeclareAcronym{DQN}{
  short = DQN ,
  long  = Deep Q Network ,
  class = abbrev
}

\DeclareAcronym{PAPR}{
  short = PAPR ,
  long  = Peak-to-Average Power Ratio ,
  class = abbrev
}

\DeclareAcronym{DNN}{
  short = DNN ,
  long  = Deep Neural Network ,
  class = abbrev
}

\DeclareAcronym{NN}{
  short = NN ,
  long  = Neural Network ,
  class = abbrev
}

\DeclareAcronym{SVM}{
  short = SVM ,
  long  = Support Vector Machine ,
  class = abbrev
}

\DeclareAcronym{CSI}{
  short = CSI ,
  long  = Channel State Information ,
  class = abbrev
}

\DeclareAcronym{CV}{
  short = CV ,
  long  = Computer Vision ,
  class = abbrev
}

\DeclareAcronym{BER}{
  short = BER ,
  long  = Bit Error Rate ,
  class = abbrev
}

\DeclareAcronym{FGSM}{
  short = FGSM ,
  long  = Fast Gradient Sign Method ,
  class = abbrev
}

\DeclareAcronym{OTA}{
  short = OTA ,
  long  = Over-the-Air ,
  class = abbrev
}

\DeclareAcronym{EW}{
  short = EW ,
  long  = Electronic Warfare ,
  class = abbrev
}

\DeclareAcronym{USRP}{
  short = USRP ,
  long  = Universal Software Radio Peripheral ,
  class = abbrev
}

\DeclareAcronym{OODA}{
  short = OODA ,
  long  = Observe Orient Decide Act ,
  class = abbrev
}

\DeclareAcronym{RFFE}{
  short = RFFE ,
  long  = Radio Frequency Front End ,
  class = abbrev
}

\DeclareAcronym{ADC}{
  short = ADC ,
  long  = Analog to Digital Converter ,
  class = abbrev
}

\DeclareAcronym{DAC}{
  short = DAC ,
  long  = Digital to Analog Converter ,
  class = abbrev
}

\DeclareAcronym{SDR}{
  short = SDR ,
  long  = Software Defined Radio ,
  class = abbrev
}

\DeclareAcronym{CR}{
  short = CR ,
  long  = Cognitive Radio ,
  class = abbrev
}

\DeclareAcronym{RF}{
  short = RF ,
  long  = Radio Frequency ,
  class = abbrev
}

\DeclareAcronym{UAP}{
  short = UAP ,
  long  = Universal Adversarial Perturbation ,
  class = abbrev
}

\DeclareAcronym{AWGN}{
  short = AWGN ,
  long  = Additive White Gaussian Noise ,
  class = abbrev
}

\DeclareAcronym{SNR}{
  short = SNR ,
  long  = Signal-to-Noise Ratio ,
  class = abbrev
}

\DeclareAcronym{DSP}{
  short = DSP ,
  long  = Digital Signal Processing ,
  class = abbrev
}

\DeclareAcronym{ML}{
  short = ML ,
  long  = Machine Learning ,
  class = abbrev
}

\DeclareAcronym{ATN}{
  short = ATN ,
  long  = Adversarial Transformation Network ,
  class = abbrev
}

\DeclareAcronym{P-ATN}{
  short = P-ATN ,
  long  = Perturbation - Adversarial Transformation Network ,
  class = abbrev
}

\DeclareAcronym{AAE}{
  short = AAE ,
  long  = Adversarial Auto-Encoder ,
  class = abbrev
}

\DeclareAcronym{ARN}{
  short = ARN ,
  long  = Adversarial Residual Network ,
  class = abbrev
}

\DeclareAcronym{LDAPM}{
  short = LDAPM ,
  long  = Linear Digital Amplitude Phase Modulation ,
  class = abbrev
}

\DeclareAcronym{QAM}{
  short = QAM ,
  long  = Quadrature Amplitude Modulation ,
  class = abbrev
}

\DeclareAcronym{PSK}{
  short = PSK ,
  long  = Phase Shift Keying ,
  class = abbrev
}

\DeclareAcronym{FSK}{
  short = FSK ,
  long  = Frequency Shift Keying ,
  class = abbrev
}

\DeclareAcronym{PAM}{
  short = PAM ,
  long  = Pulse Amplitude Modulation ,
  class = abbrev
}

\DeclareAcronym{SWaP}{
  short = SWaP ,
  long  = {Size, Weight, and Power} ,
  class = abbrev
}

\DeclareAcronym{API}{
  short = API ,
  long  = Application Programmer Interface ,
  class = abbrev
}

\DeclareAcronym{TRL}{
  short = TRL ,
  long  = Technical Readiness Level ,
  class = abbrev
}

\DeclareAcronym{PA}{
  short = PA ,
  long  = Power Amplifier ,
  class = abbrev
}

\DeclareAcronym{CFO}{
  short = CFO ,
  long  = Center Frequency Offset ,
  class = abbrev
}

\DeclareAcronym{STO}{
  short = STO ,
  long  = Sample Time Offset ,
  class = abbrev
}

\DeclareAcronym{SFO}{
  short = SFO ,
  long  = Sampling Frequency Offset ,
  class = abbrev
}

\begin{document}

\title{The RFML Ecosystem:\\ {\huge A Look at the Unique Challenges of Applying Deep Learning to Radio Frequency Applications}
}

\author{
    \IEEEauthorblockN{Lauren J. Wong\IEEEauthorrefmark{1}\IEEEauthorrefmark{3}, William H. Clark IV\IEEEauthorrefmark{1}, Bryse Flowers\IEEEauthorrefmark{2}, R. Michael Buehrer\IEEEauthorrefmark{3}, \\ Alan J. Michaels\IEEEauthorrefmark{1}, and William C. Headley\IEEEauthorrefmark{1}} \\
    \IEEEauthorblockA{\IEEEauthorrefmark{1}Hume Center for National Security and Technology, Virginia Tech} \\
    \IEEEauthorblockA{\IEEEauthorrefmark{3}Bradley Department of Electrical and Computer Engineering, Virginia Tech} \\
    \IEEEauthorblockA{\IEEEauthorrefmark{2}Department of Electrical and Computer Engineering, University of California, San Diego} \\
    \IEEEauthorblockA{Email: \{ ljwong, bill.clark, brysef, rbuehrer, ajm, cheadley \}@vt.edu}
}

\maketitle

\begin{abstract}
While deep machine learning technologies are now pervasive in state-of-the-art image recognition and natural language processing applications, only in recent years have these technologies started to sufficiently mature in applications related to wireless communications. In particular, recent research has shown deep machine learning to be an enabling technology for cognitive radio applications as well as a useful tool for supplementing expertly defined algorithms for spectrum sensing applications such as signal detection, estimation, and classification (termed here as Radio Frequency Machine Learning, or RFML). A major driver for the usage of deep machine learning in the context of wireless communications is that little, to no, \emph{a priori} knowledge of the intended spectral environment is required, given that there is an abundance of representative data to facilitate training and evaluation. However, in addition to this fundamental need for sufficient data, there are other key considerations, such as trust, security, and hardware/software issues, that must be taken into account before deploying deep machine learning systems in real-world wireless communication applications. This paper provides an overview and survey of prior work related to these major research considerations. In particular, we present their unique considerations in the RFML application space, which are not generally present in the image, audio, and/or text application spaces. 
\end{abstract}


\section{Introduction}\label{sec:intro}
\Ac{DL} has become a transformative technology for improving the capabilities of autonomous image recognition and \ac{NLP}, among many others. As a result, in recent years, \ac{DL} has been looked to in the wireless communications domain for facilitating applications such as blind spectrum sensing tasks, including signal detection, estimation, classification, and specific emitter identification, as well as an enabling technology for cognitive radio tasks such as \ac{DSA}. Given initial successes in these areas, among others, \ac{DL} is being considered as a major transformative technology in the upcoming 5G standard and is expected to be a core component of 6G technologies and beyond \cite{5g}.

\begin{figure}[t]
	\centering
	\includegraphics[width=\linewidth,trim=0 0 0 0,clip]{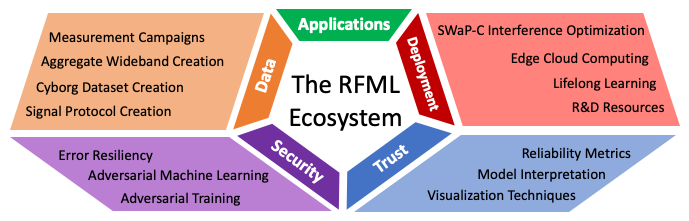}
    \caption{The \ac{RFML} ``Ecosystem" is made up of the major research thrust areas that must be considered holistically in order to utilize RFML systems in real-world applications.}
    \label{fig:ecosystem_tikz}
\end{figure}

To date, research in applying \ac{DL} to wireless communications applications, commonly termed \ac{RFML}, has predominately focused on providing novel solutions that have relied heavily on existing techniques used in other \ac{DL} modalities applied to the wireless communications application of interest. 
While this is a logical first step and has shown success, it is expected that the next great leap in \ac{RFML} capabilities will likely require uniquely tailoring deep learning architectures and techniques to the \ac{RF} domain using domain specific knowledge, just as \acp{CNN} were first designed for the unique challenges of the image processing domain.

This paper aims to facilitate this evolution by providing an overview of the key \ac{RF} domain specific considerations that must be taken into account before deploying \ac{RFML} solutions in real-world wireless communication applications. More specifically, this paper provides an overview and survey of prior works related to important research areas such as \ac{RF} dataset formulation (Section III), as well as \ac{RFML} security (Section IV), trust (Section V), and deployment (Section VI). Through this discussion, particular attention is paid to how all of these areas, the aggregate of which is here termed the \ac{RFML} ``Ecosystem" and illustrated in Figure \ref{fig:ecosystem_tikz}, interact with and affect one another as they are inextricably dependent and therefore must be considered in tandem.

The application of \ac{DL} to wireless communications applications is a very broad topic area with many different definitions and assumptions. 
For the discussions and literature review undertaken in this work, \ac{RFML} is here defined as the usage of \ac{DL} techniques that aim to reduce the amount of expert defined features and prior knowledge is needed for the intended application. 
In other words, we predominately consider works in which the input to the \ac{RFML} system is the raw \ac{RF} data \cite{rfmls}, with some discussion of prior works utilizing or deriving pre-defined expert features.

The current body of literature surveying \ac{RFML} is copious, yet narrowly focused.
More specifically, existing surveys are typically algorithm and/or application focused, and overview, compare, and contrast the learning techniques used and highlight the variety of applications and operating conditions under which \ac{RFML} is beneficial \cite{alshawaqfeh2015, jhajj2018, alsheikh2014, boutaba2018, sun2019, di2007, mao2018, clancy2007, joshi2013, erpek2020, oshea2017, chen2019, bkassiny2013, jiang2017, simeone2018}.

A few more unique surveys have also examined \ac{RFML} dataset generation using GNU Radio \cite{oshea2016datagen}, the use of big data techniques in wireless networks \cite{kibria2018bigdata}, operational considerations for using cognitive radio in a military setting \cite{tuukkanen2015military}, security and privacy challenges faced in cognitive wireless sensor networks \cite{sen2013security}, and solutions for combating practical imperfections encountered in cognitive radio system (i.e. noise uncertainty, channel/interference uncertainty, hardware imperfections, signal uncertainty, synchronization issues) \cite{sharma2015}.

In contrast, the scope of this paper is to holistically bring together these disparate works, highlighting the interdependencies, through the broader RFML ``Ecosystem" that underpins them.
 
\section{Applications}\label{sec:applications}
The \ac{RFML} ecosystem, as the name implies, is the supporting considerations in the development and deployment of \ac{RFML} applications. Therefore, before we discuss the different facets of the \ac{RFML} ecosystem, it is important to provide context through a discussion of the relevant \ac{RFML} applications found in the literature. 
An overview of the algorithms described herein, including training data types and algorithm input types, is given in Table \ref{table:app}.

\ac{ML}, and \ac{DL} in particular, has been applied to a wide variety of areas within wireless communications, including networking applications, power management, and radio control.
Given the scope of this paper, as defined above, the applications discussed here tend closer to the physical layer, and can generally be described as spectrum sensing applications.

\begin{table*}[ht]
\centering
\begin{tabular}{@{}ccccccc@{}}
\toprule
\hspace{0.5cm}\textbf{citation} & \textbf{application} & \multicolumn{3}{c}{\textbf{dataset type}} & \multicolumn{2}{c}{\textbf{input type}} \\ \midrule
 &  & \textbf{real} & \textbf{synthetic} & \textbf{augmented}\hspace{0.25cm} & \hspace{0.25cm}\textbf{raw samples} & \hspace{1cm}\textbf{features}\hspace{1cm} \\ \midrule
 \cite{Mohammed_2018} & detection & X &  &  & X &  \\ \midrule 
 \cite{Vyas_2017, oshea-detect1, Pu_2011, Yi_2018, Ge_2008, Fernandes_2018, Subekti_2018, Wunsch_2017}   &  & X &  &  &  & X \\ \midrule 
 \cite{Han_2017, Yelalwar_2018, Nawaz_2017b, Liu_2019, oshea-detect2, Chen_2015, Bixio_2009, Kim_2013, Tsakmalis_2014} &  &  & X &  &  & X \\ \midrule 
 \cite{Mody_2007} &  & X & X &  &  & X \\ \midrule 
 \cite{adv_rfml:Davaslioglu2018a, Schmidt_2017, oshea-2018-data} & classification & X &  &  & X &  \\ \midrule 
 \cite{oshea2016datagen, He_2011, hauser-amc, Hong_2017, Wu18, Li_2018, Yashashwi_2019, Oshea_2016d, west-amc2, Zhang_2018, Sang_2018, Vanhoy_2018, Clark19, wong-amc3}&  &  & X &  & X &  \\ \midrule 
  \cite{adv_rfml:Clancy2009a, Cai_2010, Reddy_2017, Kumar_2016, Mendis_2019, Bitar_2017} &  & X &  &  &  & X \\ \midrule 
 \cite{Pajarinen_2011, Nandi_1997, Fehske_2005, Ramon_2009, Popoola_2011, Abedlreheem_2012, Li_2012, Peng_2017, Mendis_2016, Nawaz_2017, Karra_2017, Hiremath_2018, Tang_2018, Ambaw_2017, Peng_2019, Jayaweera_2018, Hiremath_2019, Zhang_2017} &  &  & X &  &  & X \\ \midrule 
  \cite{vila2019} &  & X &  &  & X & X \\ \midrule 
 \cite{Kim_2003} &  & X & X &  &  & X \\ \midrule 
 \cite{Kulin_2018} &  & X & X &  & X & X \\ \midrule 
 \cite{Wang_2019, Ali_2017} &  &  & X &  & X & X \\ \midrule 
 \cite{west_2017} & \begin{tabular}[c]{@{}c@{}}detection \& \\ classification\end{tabular} & X &  &  & X &  \\ \midrule 
 \cite{Popoola_2013} &  &  & X &  &  & X \\ \midrule 
 \cite{white19, Kozy2019, Yu2020} & parameter estimation &  & X &  &  & X \\ \midrule 
 \cite{Oshea_2016, oshea2017} &  &  & X &  & X &  \\ \midrule 
 \cite{Elbakly18, AlHajri2019, Chawathe19, ts_2014} &  & X &  &  &  & X \\ \midrule 
 \cite{Oshea_2016b} & \begin{tabular}[c]{@{}c@{}}parameter estimation \& \\ classification\end{tabular} &  & X &  & X &  \\ \midrule 
\cite{merchant_2018, merchant_2019a, merchant_2019b} & SEI &  &  & X &  & X \\ \midrule 
\cite{wong-sei1} &  & X &  &  & X &  \\ \midrule  
\cite{wong-sei2, McGinthy} &  &  & X &  & X &  \\ \midrule 
\cite{Tandiya18} & Anomaly Detection &  & X &  &  & X \\ \midrule 
\cite{oshea_2016e} &  &  & X &  & X &  \\ \midrule 
\cite{Chang19, xu-dsa, xu2020} & DSA &  & X &  &  & X \\ \midrule 
\cite{Kirk19, Selvi20, Thornton20, Martone20} & Cognitive Radar &  & X &  &  & X \\ \midrule 
\cite{oshea_2018} & Cognitive Radio & X &  &  & X &  \\ \midrule
\cite{dorner_2018} &  & X & X &  & X &  \\\midrule
\cite{erpek_2018, oshea_2017} &  &  & X &  & X &  \\ \midrule 
\cite{oshea_2016f} &  &  & X &  & X & X \\ \bottomrule
 
\end{tabular}%
\caption{An overview of the dataset and input types used in popular \ac{RFML} applications.}
\label{table:app}
\end{table*}

\subsection{Spectrum Sensing}

Spectrum sensing is the process of gaining knowledge of a given spectral environment with little, to no, \emph{a priori} knowledge of the environment. 
Spectrum sensing is primarily made up of the following \ac{DSP} tasks: signal detection \cite{Thilina_2013, Vyas_2017, Han_2017, Yelalwar_2018, Nawaz_2017b, Mohammed_2018, Liu_2019, west_2017, oshea-detect1, oshea-detect2, Popoola_2013, Pu_2011, Chen_2015, Yi_2018, Mody_2007, Ge_2008, Bixio_2009, Kim_2013, Tsakmalis_2014, Fernandes_2018, Subekti_2018, Wunsch_2017, white19}, signal parameter estimation \cite{Oshea_2016, Oshea_2016b, oshea2017, Elbakly18, AlHajri2019, Chawathe19, ts_2014}, signal classification \cite{adv_rfml:Clancy2009a,oshea2016datagen, west_2017, adv_rfml:Davaslioglu2018a, Pajarinen_2011, Nandi_1997, Kim_2003, Fehske_2005, Ramon_2009, Popoola_2011, Cai_2010, He_2011, Abedlreheem_2012, Li_2012, Peng_2017, Reddy_2017, Schmidt_2017, Kumar_2016, Mendis_2016, Nawaz_2017, Oshea_2016b, Karra_2017, hauser-amc, Mendis_2019, Bitar_2017, Hiremath_2018, Tang_2018, Ambaw_2017, Hong_2017, Kulin_2018, Wu18, Li_2018, Yashashwi_2019, Peng_2019, Wang_2019, Jayaweera_2018, Hiremath_2019, Popoola_2013, Ali_2017, Oshea_2016d, west-amc2, oshea-2018-data, Zhang_2018, Sang_2018, Vanhoy_2018, Wang_ARL_2019, Clark19, Zhang_2017, wong-amc3}, emitter identification/fingerprinting \cite{merchant_2018, wong-sei1, wong-sei2, merchant_2019a, merchant_2019b, McGinthy}, and anomaly detection \cite{Tandiya18, oshea_2016e}.
These spectrum sensing tasks are of fundamental importance in both military and commercial applications. For example, in military communications, spectrum sensing is critically important for jamming/anti-jamming, eavesdropping, localization, and demodulation of adversary communications \cite{jemso2020}. In commercial communications, spectrum sensing is the primary enabler of \ac{DSA} in which spectrum users collaboratively utilize spectrum resources without the need for strict, and typically inefficient, spectrum licenses or centralized spectrum managers \cite{Chang19,xu-dsa,xu2020, Kirk19,Selvi20,Thornton20,Martone20}. 
Finally, as the quantity of wireless devices continue to grow and rest in the hands of the general public, localization can be a significant tool for a multitude of emergency and safety applications \cite{Kozy2019, Yu2020}, such as search and rescue operations.

\subsubsection{\ac{AMC}} One of the earliest, and perhaps the most researched, applications of \ac{RFML} for spectrum sensing is in the area of modulation classification, likely due to the fact machine learning techniques have historically performed extremely well on classification tasks. 
Traditional modulation classification techniques typically consist of two signal processing stages: feature extraction and pattern recognition. 
Traditionally, the feature extraction stage has relied on the use of so-called ``expert features" in which a human domain-expert pre-defines a set of signal features that allow for statistical separation of the modulation classes of interest, examples of which can be found in \cite{4167652}. 
These expertly defined signal features are extracted from the raw received signal during a potentially time intensive and computationally expensive pre-processing stage. 
These expert features are then used as input to a pattern recognition algorithm, which might consist of decision trees, support vector machines, feed-forward neural networks, among many others. 

\ac{RFML} based approaches have aimed to replace the human intelligence and domain expertise required to identify and characterize these features using deep neural networks and advanced architectures, such as \acp{CNN} and \acp{RNN}, to both blindly and automatically identify separating features and classify signals of interest, with minimal pre-processing and less \emph{a priori} knowledge \cite{west-amc2, Wu18, Clark19, wong-amc3, vila2019}. Given the significant research in \ac{RFML}-based modulation classification, it can be argued that \ac{AMC} is one of the most mature fields in \ac{RFML}, and has been deployed in real-world products \cite{signaleye}.

\subsubsection{Signal Detection}
Another spectrum sensing area seeing a particular increase in the \ac{RFML} literature is signal detection \cite{oshea-detect1, oshea-detect2,white19}. 
This is one key example in which image processing techniques have been directly applied to solve \ac{RFML} problems.
For example, in \cite{oshea-detect1, oshea-detect2,white19}, the raw \ac{IQ} samples were converted into spectrum waterfall plots, where the spectrum information in the time-frequency plane is viewed as an image.
This has allowed a rich class of image processing techniques to be applied directly to signal detection applications for positive \ac{SNR} environments. 


\subsubsection{Specific Emitter Identification}
\ac{SEI}, also known as \ac{RF} Fingerprinting, is an application that has benefited greatly from the advent of \ac{RFML} \cite{wong-sei1, wong-sei2, oshea-amc1, McGinthy, Polak}. 
For the purposes of \ac{SEI}, classification is aimed not towards the transmitted signal, but the transmitter itself, which is possible due to slight (but consistent) differences between emitters, such as \ac{IQ} imbalances, amplifier non-idealities, and other imperfections caused during the manufacturing process \cite{wong-sei2}, as well as geographical differences such as propagation channels and angle of arrival \cite{neu_tutorial}.

These differences not only exist between transmitter brands and models, but amongst transmitters of the \textit{same} brand and model, which may have been manufactured side-by-side.
Due to the difficult nature of defining expert features to distinguish between these subtle differences, the usage of DL within SEI has shown great benefits.
More specifically, given the vast number of existing devices, each of which exhibit nearly imperceptible differences, it is near impossible to accurately predict, model, and extract discriminating features. 
Given these limitations, \ac{DL}-based solutions, more specifically \acp{CNN}, which ingest raw \ac{RF} data, have been used in order to learn the discriminating features for identifying transmitters \cite{wong-sei1, wong-sei2}.

\subsubsection{Parameter Estimation}
Parameter estimation is the process of determining relevant discriminating features of a transmitted signal (center frequency, bandwidth, duration, etc) or its propagation channel (time/frequency drifts, shadowing loss, multi-path taps, etc). 
Often, parameter estimation in \ac{RFML} is a byproduct of other spectrum sensing tasks.
For example, for signal detection applications utilizing spectral waterfall images as input, estimates of the bandwidth, center frequency, and duration are found during the detection process. 
However, applying \ac{RFML} directly for parameter estimation has also shown promise, especially in the area of localization. 

Localization is incredibly important in both military and commercial communications. 
Although, traditionally, localization techniques have relied on expert defined features such as received signal strength \cite{Bacak}, in recent years a more rich set of \ac{RF} measurements have been used \cite{Bacak, AlHajri2019}.  
In recent work, channel state information is used to reach cm-level accuracy \cite{Yu2020} and tackle the difficult problem of indoor 3D (i.e. multi-floor) localization \cite{Elbakly18}.  
There has also been progress towards classifying indoor locations using RF samples \cite{Chawathe19}.

\subsubsection{Anomaly Detection}
Finally, an emerging \ac{RFML} application area is anomalous event detection. 
In these applications, the intent is for a deep learning approach to learn the baseline environment and detect/classify deviations from this baseline (so-called anomalies). 
Such applications may also be used to detect adversarial attacks or to identify out-of-distribution examples, as discussed further in Sections \ref{sec:security} and \ref{sec:trust}.

An example of this budding area of research can be found in \cite{Tandiya18}, where RF spectrum activities are monitored and analyzed using a deep predictive coding \acp{NN} to identify anomalous wireless emissions within spectrograms.
Similarly, in \cite{oshea_2016e}, the authors utilized recurrent neural predictive models to identify anomalies in raw IQ data.

\subsection{Cognitive Radio}
In the literature, the terms \ac{RFML} and \emph{cognitive radio} have often been mistakenly used interchangeably.
In comparison to the definition of \ac{RFML} given in Section \ref{sec:intro}, cognitive radio is the use of software-defined radios coupled with state-of-the-art \ac{RFML} techniques to enable radios to intelligently, and efficiently, utilize its hardware and spectral resources \cite{oshea_2016f}. 
\ac{RFML}-based solutions have further enabled the realization of cognitive radio through two keys areas, namely through improved spectrum sensing capabilities (as just discussed) and through direct replacement of fixed signal processing stages with \ac{RFML}-enabled dynamic stages. 

The primary goal of cognitive radio is to adapt to changing channel conditions without needing a human in the loop or time intensive re-configurations. 
While the usage of \ac{RFML}-based techniques for supplementing traditional \ac{DSP} techniques in cognitive radios is still in its infancy, these initial works show promise. 
One intriguing example of cognitive radio is the use of \ac{GAN} or \ac{AE} for end-to-end communications systems design.  
More specifically, in \cite{oshea2017, oshea_2018, erpek_2018, oshea_2017, dorner_2018}, the authors developed algorithms that allow the neural network model to design the physical layer communications protocol (including modulation scheme, coding scheme, filtering, etc.), using bit or symbol error rate as the performance metric, to overcome the challenge of choosing an appropriate modulation scheme for an unknown, uncharacterized, or changing channel. 
These approaches have shown promise for both \ac{SISO} and \ac{MIMO} radio communications.

\section{Dataset Creation}\label{sec:data}
In any application of \ac{DL}, representative and well-labeled datasets are of critical importance for training and/or evaluation. Datasets can generally be categorized into one of three types, namely simulated, captured/collected, and augmented.
Simulated datasets refer to synthetically generated data, in which the transmitter, channel, and receiver are all modeled in software/hardware. 
In contrast, captured datasets are collected from signals that have been transmitted over a wireless channel. Finally, augmented datasets combine simulated and captured data by adding synthetic perturbations to captured data and/or adding synthetic signals to captured channels.

Simulated datasets are the most commonly used in current \ac{RFML} literature as they are the most straightforward to compile and label using publicly available toolsets such as GNU Radio \cite{clark_gnu}, liquid-dsp \cite{liquid}, and MATLAB \cite{matlab} among others, and therefore lends itself well to initial development
\cite{Nandi_1997,Kim_2003,Fehske_2005,Mody_2007,Ge_2008,Bixio_2009,adv_rfml:Clancy2009a,Ramon_2009,Popoola_2011,Kang_2011,Pu_2011,He_2011,Abedlreheem_2012,Li_2012,Thilina_2013,Kim_2013,Tsakmalis_2014,Chen_2015,Peng_2017,Reddy_2017,Mendis_2016,Nawaz_2017,Oshea_2016,Oshea_2016c,Karra_2017,Yelalwar_2018,hauser-amc,Hiremath_2018,Nawaz_2017b,Tang_2018,Ambaw_2017,Hong_2017,Kulin_2018,Wu18,Li_2018,Tandiya_2018,Yashashwi_2019,Peng_2019,adv_rfml:Davaslioglu2018a,Wang_2019,Liu_2019,Subekti_2018,Jayaweera_2018,Popoola_2013,Ali_2017,oshea2016datagen,Oshea_2016d,west-amc2,oshea-2018-data,Zhang_2018,adv_rfml:Sadeghi2018a,Sang_2018,west_2017,Zheng_2019,Vanhoy_2018,Shapero_2018,Wang_ARL_2019,Clark19,oshea-detect2, wong-sei2, merchant_2019b}.
Unlike in image processing \cite{synth_images}, the same equations and processes used to transmit waveforms in real systems can be used directly in simulation, due to their man-made nature \cite{oshea2016datagen}. 
Additionally, for simplistic environments, mathematical models can be used to reasonably describe common degradations such as additive interference, channel effects, and transceiver imperfections. 
As a result, synthetically generated \ac{RFML} datasets can be good analogs for captured \ac{RFML} datasets, if carefully crafted and known models exist.

Captured data is necessary for test and evaluation prior to real-world deployment, but requires significant man-hours and resources to properly collect \cite{Mody_2007, Ge_2008, Cai_2010, Pu_2011,Reddy_2017,Schmidt_2017,Kumar_2016,Vyas_2017,Fernandes_2018,Mendis_2019,merchant_2018,Bitar_2017,Kulin_2018,Testi_2018,wong-sei1,Mohammed_2018,Hiremath_2019,Popoola_2013,oshea-2018-data,Yi_2018,sankhe2019oracle,oshea-detect1, oshea-detect2, Elbakly18, AlHajri2019, Chawathe19, ts_2014, merchant_2019a}. 
This type of data is necessary, because a capture from the real environment will include all of the different degradation types that are of concern in practical RF situations, some of which may be missing from a simulated dataset due to inaccurate modeling; captured data often reduces end user resistance and doubt surrounding the system. 
As just mentioned, the downside to utilizing capture data are the man-hours and resources required to gather sufficiently diverse captures for producing a training and/or evaluation datasets, and then to label it correctly \cite{hauser-amc}. 
This is the fundamental reason that augmented datasets are used, which combine simulated and captured data to increase the quantity of data available for training, or to create more realistic datasets \cite{Mody_2007,adv_rfml:Davaslioglu2018a,Schmidt_2017,merchant_2018,Kulin_2018,Wang_ARL_2019}.

Finally, while not as optimal as real world data, augmented datasets aim to provide a ``best of both worlds" approach by minimizing the limitations of synthetic datasets (i.e. real-world model accuracy) and reducing the amount of captured data needed. At its most basic, augmented datasets shuffle a small subset of real-world data into a larger synthetic dataset. The intent with this approach is to utilize the synthetic data to teach the \ac{DL} model the features and characteristics of signals that can be well modeled in software (modulation schemes, simple channel models, etc), while the captured data is used to teach the \ac{DL} model the features and characteristics of signals that cannot be modeled well (i.e. transmitter/receiver imperfections).
A more complex augmented dataset might include injecting synthetic waveforms into captured spectrum. 
Such a dataset would be useful in testing detection and classification performance of the signals in a congested or interference environment with real-world transmitted signals. 
As another example, synthetic noise could be added to real world captures to decrease the \ac{SNR} without performing additional signal captures, and increasing the range of test \acp{SNR}.
However, there are a multitude of open research questions in the development and usage of augmented datasets. 
While many of these questions are application and environment specific, perhaps the most important of these open questions is how to balance the amount of real, synthetic, and augmented data used in training datasets to avoid network bias. 

\subsection{Real World Considerations}\label{sec:data-rwc}
No matter the type of dataset being created, there are a number of real-world effects that need to be considered, as they can dramatically impact \ac{RFML} system performance.
When measuring the performance of an \ac{RFML} device on laboratory-measured or synthetic data versus observed data, the primary difference is often that the trained environment is pristine in comparison to a real environment.
This is due to the fact that signals that have propagated in the physical world undergo degradation from multiple overlapping sources not typically encountered in a laboratory.
More specifically, both the transmitter and receiver will cause some distortion in the form of non-linearities, additive noise, timing offsets, frequency offsets, phase offsets, sample rate mismatches, and/or amplitude offsets, all potentially time varying, altering the signal from the ideal.
Additionally, the physical medium, or channel, through which the signal propagates can change as the transceivers travel around or the environment shifts, allowing delayed imperfect reflections to overlap with the direct path causing time and frequency varying interference with itself.

Depending on the application, the distortion to the waveform caused by the transmitter in particular may be a parameter of interest or may be considered a \emph{nuisance parameter} requiring an ensemble of emitters to model an \textit{average} emitter.
For example, applications such as \ac{SEI} depend upon transmitter imperfections to distinguish between emitters.
Meanwhile, within \ac{AMC} transmitter imperfections are considered nuisance parameters, as the goal of \ac{AMC} is to identify the modulation class, regardless of the emitter.
In the case of receiver distortions \cite{hauser-amc,Bacak}, we find that natural reception variations such as sampling rate differentials, frequency offsets, and varying \ac{SNR} lead to the requirement for generalized training across each of the parameters \cite{Adesina}.

An important note for these hardware variations is that high quality hardware tends to have fewer distortions. 
For example, military transmitters are often harder to \ac{RF} fingerprint than low-cost \ac{IoT} transmitters, and \ac{RF} signal behaviors learned through a high-quality receiver will translate more easily to another high quality receiver rather than one of lower quality. 
Moreover, the non-linearities that contribute to these variations are often dependent upon technology and hardware configurations, and lack of synchronization between devices will exacerbate the distortion caused by both the physical medium and the devices.
In this case detection and isolation routines are needed to select spectrum of interest, which in turn will introduce measurement errors in time, frequency, and phase offsets between the transceivers. These measurement errors should also be modeled in order create a realistic simulated dataset.

A second area of real-world consideration for \ac{RFML} system performance is the signal propagation and/or channel effects.
The baseline laboratory training environment for virtually all \ac{RF} systems is an \ac{AWGN} channel, yet real-world channels have time-varying, often colored spectra, and uncontrolled \ac{RF} interference sources such as other signals, impulsive noise (i.e. lightning), and non-linear effects associated with bursty packet transmissions.
Relative motion between platforms, co-channel/adjacent channel interference, and multi-path should also be considered, yet is often ignored during laboratory training.

Finally, a third consideration, and we believe largely an open problem as we look to scale up the deployment of \ac{RFML} solutions in real-world scenarios is the use of \textit{transfer learning} \cite{Wagle1,Wagle2}, where the behaviors learned and observations collected can be shared between sensors, as well as \textit{online or incremental learning}, where the behaviors learned are modified over time as a function of a changing environment.
For example, automated vehicles could benefit greatly from sharing their observations with the neighboring platforms while operating in a \ac{V2V}/\ac{V2I} environment.
However, abrupt differences in urban environments and military signal collection/analysis platforms (e.g., ship or \ac{UAV}) make this challenging.
Any future works must consider that the behaviors learned on one platform will be influenced by their \ac{RF} hardware, which is distinct and possibly vastly different from a second platform.
Moreover, the acceptance of online, incremental, and transfer learning poses the significant risks that any learned or transferred behaviors may misrepresent the actual environment (unintentionally or through adversarial interaction), suggesting the need for periodic \textit{save} points for all relevant parameters.

\subsection{Developing a Dataset}
Taking into account the three types of datasets (simulated, captured, and augmented), and some of the concerns relevant when operating an \ac{RFML} system outside of a laboratory setting, what follows are general guidelines that should be observed when creating a new \ac{RFML} dataset.
The first step, no matter the type of dataset being created or intended application, is identifying the expected degradations in the deployed environment (i.e. channel types, transmitter imperfections, \ac{SNR}, etc) and categorizing whether each potential degradation is fundamental to the application.
Minimally, each degradation considered fundamental to the application should be described by either
\begin{itemize}
\item defining (mathematically) how the degradation is applied, in simulated and augmented datasets, or
\item defining the conditions under which the degradation is collected (i.e. hardware used, environment, etc), for captured datasets.
\end{itemize}
For each identified nuisance degradation, attempts should be made to generalize over observations of the degradation, such as sweeping over the impairment range in simulations or changing the transmission devices or environment in some way while capturing the dataset.

During the dataset creation process, whether through simulation or collection, correctly and completely labelling the data is of the utmost importance.
Ideally, though not practically, every parameter should be recorded as metadata associated with the observations in the dataset to increase the number of applications pertinent to the dataset \cite{sigmf}.
Qualitative descriptions can also be used to provide as much description as is feasible.
Minimally, the parameter of interest to the application (the modulation class in the case of \ac{AMC}, for example) should be recorded. 
However, the value of generating and providing datasets with significant diversity, documentation, and open usage rights should not be lost on the community as the gains observed in the image processing domain were realized with the help of crowd sourcing efforts and donations of people's time \cite{imagenet}.

\subsection{Data Used in Existing Works}

A non-exhaustive search for publicly available datasets within \ac{RFML} turns up datasets released by Geotec \cite{geotec_datasets} for Emitter Localization, DeepSig \cite{deepsig_datasets} for \ac{AMC}, and by Genesys at Northeastern University \cite{genesys_datasets} for \ac{RF} Fingerprinting.
These published datasets were generated for and used in original published works \cite{ts_2014,oshea2016datagen,oshea-2018-data,sankhe2019oracle}, and have created a common point of comparison for different ML approaches within the literature. 
Though establishing whether these specific datasets can be trusted is outside the scope of the work, knowledge of how the signals in any \ac{RF} dataset were generated and understanding how to extend/modify said dataset to suit one's needs is critical in answering whether that dataset can be trusted.
Otherwise every signal (and the associated metadata, if applicable) should go through some form of validation by the user, prior to accepting that dataset, but such a validation process is often prohibitive both in computation and time.

\begin{figure}[t]
	\centering
	\includegraphics[width=0.95\linewidth,trim=0 0 0 0,clip]{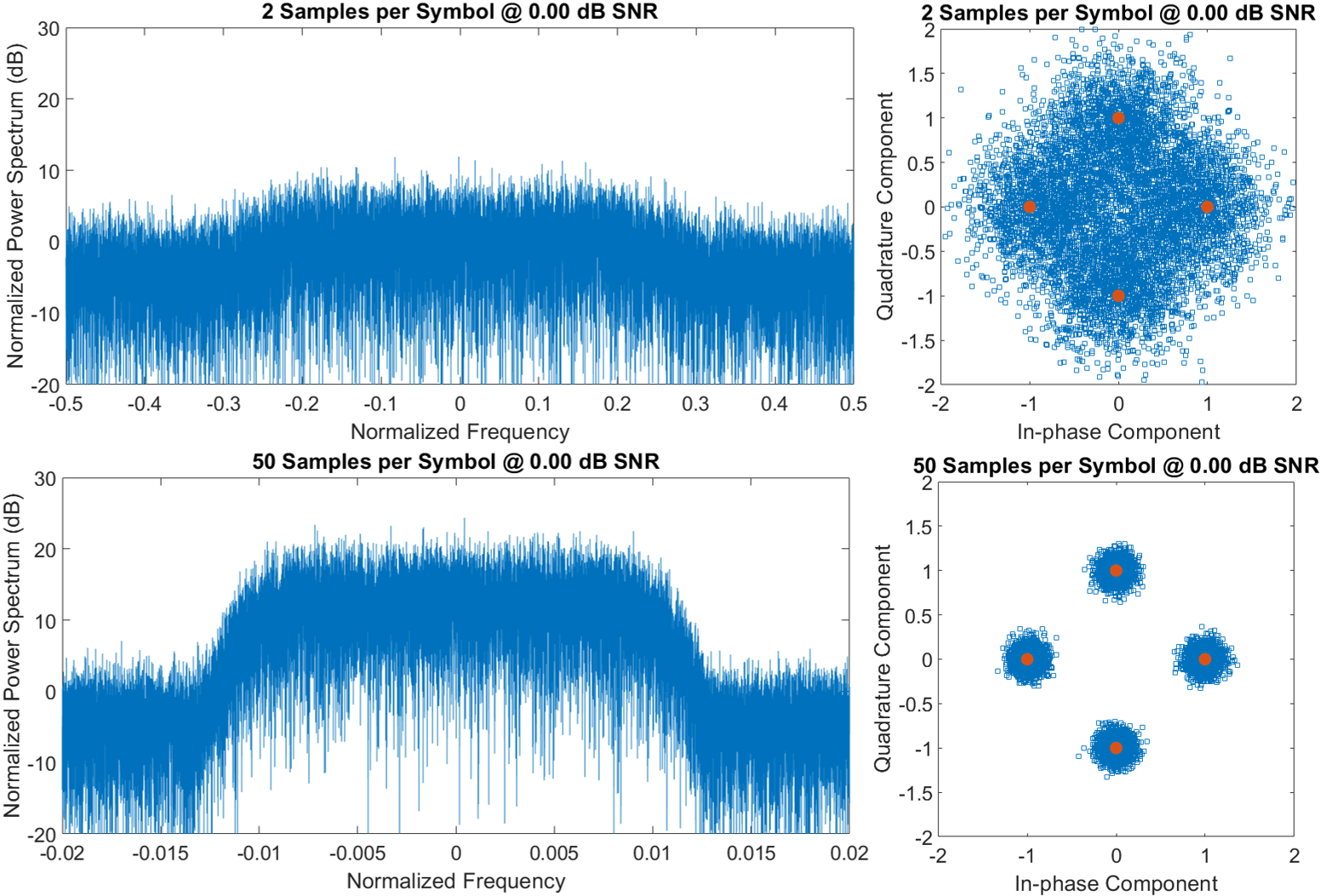}
	\caption{An example of the difficulties of direct comparison when the dataset's parameters are not explicitly defined. In this case, both signals can claim an SNR value of 0 dB, but the second is significantly oversampled and allows for either preprocessing or learning a filter-like behavior raising the apparent SNR observed during processing.}
	\label{fig:snr_caution}
\end{figure}

Given the limited availability of publicly available \ac{RFML} datasets, the majority of existing works create their own.
For these works and future works, it is critical to describe the parameter space from which the data was generated in the final publication, so that future researchers can reproduce the results.
As an example, we consider the signal shown in Figure \ref{fig:snr_caution}, where two signals have been generated with the same \ac{SNR} (but with vastly different sampling rates), as defined by
\begin{equation}\label{eq:snr}
    \Gamma_{dB} = 10\log_{10}\left(\sum_{n\in N}\left|s_n\right|^2\right) - 10\log_{10}\left(\sum_{n\in N}\left|\nu_n\right|^2\right)
\end{equation}
where $s_n$ is the $N$ samples of the signal of interest and $\nu_n$ is the $N$ samples of \ac{AWGN} and $\Gamma_{dB}$ is the \ac{SNR} value in dB. 
Traditional \ac{DSP} dictates that the bottom signal can achieve a higher maximum \ac{SNR} using a matched filter, as is evident in the constellation plots.
As most \ac{RFML} applications describe performance as a function \ac{SNR}, not including parameters such as sampling rates can greatly impede the ability to reproduce results and can therefore lead to false comparisons in subsequent publications. 

Further exacerbating this issue, results could also be described in reference to the energy per bit or $E_b/N_0$, further creating ambiguity. In this case, the effect of sample rate scaling is already accounted for, but this definition suffers when considering waveforms with no direct definition of $E_b/N_0$.

\section{Security}\label{sec:security}
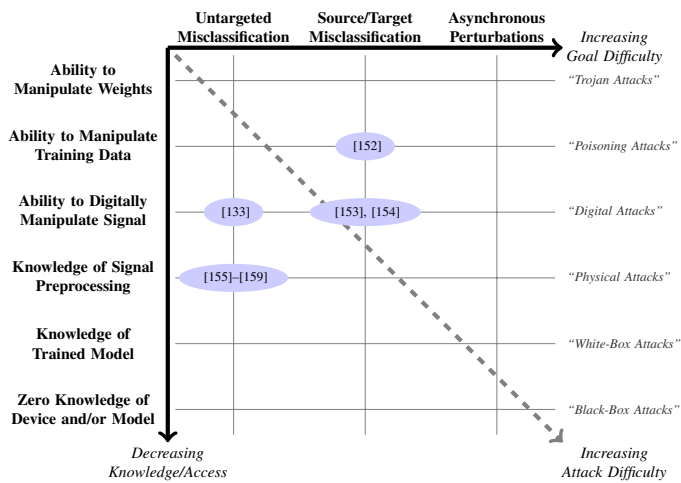
\begin{figure}
\centering

\begin{tikzpicture}[related work/.style={scale=0.5, shape=ellipse, fill=blue!20}, label/.style={scale=0.6, font=\bfseries, align=center, text width=3.5cm}, axes annotation/.style={scale=0.6, font=\itshape, align=center, text width=3.5cm}, watermark/.style={scale=0.5, font=\itshape, align=center, shape=rectangle, fill=white, opacity=0.8}]

    \pgfmathsetmacro{\HSPACING}{1.75}
    \pgfmathsetmacro{\VSPACING}{0.875}
    \pgfmathsetmacro{\NRIGHT}{3}
    \pgfmathsetmacro{\NDOWN}{6}

    \pgfmathsetmacro{\blx}{\HSPACING / 2}
    \pgfmathsetmacro{\bly}{\VSPACING / 2}
    \coordinate (bl) at (\blx, \bly);
    \pgfmathsetmacro{\tlx}{\HSPACING / 2}
    \pgfmathsetmacro{\tly}{\NDOWN * \VSPACING + \VSPACING / 2}
    \coordinate (tl) at (\tlx, \tly);
    \pgfmathsetmacro{\trx}{\NRIGHT * \HSPACING + \HSPACING / 2}
    \pgfmathsetmacro{\try}{\NDOWN * \VSPACING + \VSPACING / 2}
    \coordinate (tr) at (\trx, \try);
    \pgfmathsetmacro{\brx}{\NRIGHT * \HSPACING + \HSPACING / 2}
    \pgfmathsetmacro{\bry}{\VSPACING / 2}
    \coordinate (br) at (\brx, \bry);

    \coordinate (x) at (1, 0);
    \coordinate (y) at (0, 1);
    
    \pgfmathsetmacro{\centerx}{\blx + (\brx - \blx) / 2}

    \coordinate (hoffset) at (0.1, 0.0);
    \coordinate (voffset) at (0.0, 0.1);

    \draw[ultra thick, ->] (tl) -- (tr); 
    \draw[ultra thick, ->] (tl) -- (bl); 
    \draw[ultra thick, dashed, ->, gray] ($ (tl) + (hoffset) - (voffset) $) -- (br); 
    \draw[xstep=\HSPACING, ystep=\VSPACING, gray, very thin] ($ (bl) + (hoffset) + (voffset) $) grid ($ (tr) - (hoffset) - (voffset) $);
    
    \node[style=axes annotation, right] at (\trx - \HSPACING / 4, \try) {Increasing Goal Difficulty};
    \node[style=axes annotation, below] at (bl) {Decreasing Knowledge/Access};
    \node[style=axes annotation, below right] at (\brx - \HSPACING / 4, \bry) {Increasing Attack Difficulty};

    \node[style=label, above] at (\tlx + 1 * \HSPACING - \HSPACING / 2, \tly) {Untargeted Misclassification};
    \node[style=label, above] at (\tlx + 2 * \HSPACING - \HSPACING / 2, \tly) {Source/Target Misclassification};
    \node[style=label, above] at (\tlx + 3 * \HSPACING - \HSPACING / 2, \tly) {Asynchronous Perturbations};

    \node[style=label, left] at (\tlx, \tly - 1 * \VSPACING + \VSPACING / 2) {Ability to Manipulate Weights};
    \node[style=label, left] at (\tlx, \tly - 2 * \VSPACING + \VSPACING / 2) {Ability to Manipulate Training Data};
    \node[style=label, left] at (\tlx, \tly - 3 * \VSPACING + \VSPACING / 2) {Ability to Digitally Manipulate Signal};
    \node[style=label, left] at (\tlx, \tly - 4 * \VSPACING + \VSPACING / 2) {Knowledge of Signal Preprocessing};
    \node[style=label, left] at (\tlx, \tly - 5 * \VSPACING + \VSPACING / 2) {Knowledge of Trained Model};
    \node[style=label, left] at (\tlx, \tly - 6 * \VSPACING + \VSPACING / 2) {Zero Knowledge of Device and/or Model};

    \node[style=watermark, right] at (\trx, \tly - 1 * \VSPACING + \VSPACING / 2) {``Trojan Attacks''};
    \node[style=watermark, right] at (\trx, \tly - 2 * \VSPACING + \VSPACING / 2) {``Poisoning Attacks''};
    \node[style=watermark, right] at (\trx, \tly - 3 * \VSPACING + \VSPACING / 2) {``Digital Attacks''};
    \node[style=watermark, right] at (\trx, \tly - 4 * \VSPACING + \VSPACING / 2) {``Physical Attacks''};
    \node[style=watermark, right] at (\trx, \tly - 5 * \VSPACING + \VSPACING / 2) {``White-Box Attacks''};
    \node[style=watermark, right] at (\trx, \tly - 6 * \VSPACING + \VSPACING / 2) {``Black-Box Attacks''};

    \node[style=related work] at (2 * \HSPACING, 5 * \VSPACING) {\cite{adv_rfml:Davaslioglu2019a}};

    \node[style=related work] at (1 * \HSPACING, 4 * \VSPACING) {\cite{adv_rfml:Sadeghi2018a}};

    \node[style=related work] at (2 * \HSPACING, 4 * \VSPACING) {\cite{adv_rfml:Bair2019a, adv_rfml:Kokalj-Filipovic2019b}};
    
    \node[style=related work] at (1 * \HSPACING, 3 * \VSPACING) {\cite{adv_rfml:Flowers2019a, adv_rfml:Flowers2020a, adv_rfml:Hameed2019a, adv_rfml:Sadeghi2019a, adv_rfml:Usama2019a}};
\end{tikzpicture}

    \caption{Threat Model for \ac{RFML} adopted from \cite{adv:Papernot2016a, adv_rfml:Flowers2020a} and including related work.}
    \label{fig:threat_model}
\end{figure}

While \acs{AI/ML} has been adopted in some format in nearly all industries in recent years, it's limitations in adversarial settings have been well documented in modalities such as \ac{CV} \cite{adv:Akhtar2018a}, audio recognition \cite{adv:Qin2019a, adv:Carlini2018a, adv:Taori2019a}, and \ac{NLP} \cite{adv:Zhang2020a}.
While the attacks demonstrated in other modalities serve as a prescient warning for applications of \ac{RFML} and many parallels can be drawn, recent work has shown there are unique considerations for \acs{AI} security in the context of \acs{RF} due to the nature of wireless propagation, pre-processing steps before \ac{RF} signals of interest are input to \acp{DNN}, and the fact that wireless communications are generally quite sensitive to perturbations in the transmission.
Therefore, while this section provides a brief overview of \ac{AI} security in general, the focus is on the unique considerations in \ac{RFML} for which a Threat Model is provided in Figure \ref{fig:threat_model} which quickly categorizes the related work in the area.

\subsection{Adversarial Machine Learning} \label{ss:adv}
When discussing \ac{AI} security, the conversation primarily revolves around Adversarial \acs{ML} which concerns the development of algorithms to attack data driven models and to defend against such attacks.
Discussions of Adversarial \ac{ML} \cite{adv:Chakraborty2018a, adv:Huang2011a} date back at least 15 years \cite{adv:Barreno2006a, adv:Barreno2010a, adv:Biggio2018a} and have broadened to include exploratory attacks that seek to learn information about (or replicate) the classifier \cite{adv:e2016a} or training data \cite{adv:Shokri2017a} through limited probes on the model to observe it's input/output relationship.
However, the most recent explosion in concern for the vulnerabilities of \acp{DNN} specifically is largely credited to the \ac{FGSM} attack \cite{adv:Goodfellow2015a} which showed that \ac{CV} models are vulnerable to small, human imperceptible, perturbations to their input images that cause a misclassification -- and these perturbations are relatively computationally cheap to create, requiring only a single back-propagation instead of a costly optimization loop.
This manipulation of the model's inputs to achieve a goal such as misclassification (not classified as the true class of the input) or targeted misclassification (forcing the model to classify the input as a \emph{specific} and \emph{different} class) is termed an evasion attack and is the most widely studied sub-field of Adversarial \ac{ML}, including in \ac{RFML}.

It should also be noted that while this section focuses primarily on attacks on signal classification, the adversarial attacks can be more broadly applied to other \ac{RFML} tasks.
Generally speaking, an evasion attack influences the input of a \ac{DNN} to change its output.
Therefore, any application of \acp{DNN} for spectrum sensing or cognitive radio discussed in Section \ref{sec:applications} is susceptible to attack.

\subsubsection{Evasion} \label{sss:adv:evasion}
Evasion attacks are most prevalent in the study of classification tasks where a key constraint in these attacks is to ensure they remain \emph{imperceptible} to the intended receiver, which is uniquely defined in the context of wireless.
Evasion attacks can further be categorized as untargeted or targeted digital attacks, as will be discussed further below.


\paragraph{Untargeted Digital Attacks}
Untargeted digital attacks can be defined as evasion attacks in which the goal is solely misclassification.
\ac{RFML} models have been shown to be just as vulnerable to these untargeted adversarial attacks as their \ac{DNN} counterparts in \ac{CV}.
Specifically, both \cite{adv_rfml:Sadeghi2018a} and \cite{adv_rfml:Flowers2020a} showed that the \ac{FGSM} attack is sufficient to completely evade \ac{AMC} by a \ac{DNN} with a perturbation that is 10 dB below the actual signal.
While \ac{FGSM} is a computationally cheap method for creating adversarial examples, the large body of literature in adversarial \ac{ML} for \ac{CV} has yielded algorithms that can evade classifiers with even smaller perturbations.
In \cite{adv_rfml:Usama2019a}, a more sophisticated adversarial methodology \cite{adv:Carlini2017a} was used to carry out an attack on \ac{AMC}; not only was this attack successful for a \ac{DNN}, but, when the adversarial examples were input to classifiers not based on \acp{DNN} (e.g. \ac{SVM}, Decision Trees, Random Forests, etc.) the models had similar decreases in accuracy.
Therefore, although adversarial \ac{ML} methodologies use \acp{DNN} to craft adversarial examples due to the need for back propagation, they are transferable across various classification methodologies.
Thus, the perturbations cannot simply be noise specific to a \ac{DNN} model, they must be changing something inherent to the signal properties that are used by many methodologies for classification.

\paragraph{Targeted Digital Attacks}
As previously mentioned, the goal of targeted digital attacks is to force a model to make a \emph{specific} misclassification.
By more closely examining \emph{how} \ac{AMC} \acp{DNN} break down under evasion attacks, other work has also shown that Adversarial \ac{ML} takes advantage of something inherent to the properties of man-made signals.
More specifically, modulation formats for wireless communications are man made; thus, they can be intuitively grouped into a hierarchical structure.
For instance, analog modulations, such as the Amplitude Modulation and Frequency Modulation used in older vehicle radios, are distinctly separate from digital modulations used to carry discrete symbols representing the bits of a data transmission.
Within digital modulations, the formats can be hierarchically grouped into whether they represent symbols in the frequency domain (\ac{FSK}), in the signal's phase (\ac{PSK}), or in both the signal's phase and amplitude (\ac{QAM}).
One would expect that a \ac{DNN} would learn this intuitive grouping as well and therefore it would more easily confuse an analog modulation with another analog modulation than it would mislabel an analog modulation as a digital transmission.
In fact, \cite{adv_rfml:Bair2019a} used the Momentum Iterative \ac{FGSM} \cite{adv:Dong2018a} attack to show that this is precisely the case.
The authors showed that higher power adversarial perturbations are required to move between source/target pairs belonging to different coarse-grained categories than to target a class belonging to the same category as the source signal.

\paragraph{Rubbish Class Examples/Fooling Images}
Other research has considered the ability to create examples that are classified as some target class but have no semantic meaning, commonly referred to as \emph{Rubbish Class Examples} \cite{adv:Goodfellow2015a}, \emph{Fooling Images} \cite{adv:Nguyen2015a}, or, in the context of wireless communications, \emph{Spoofing Attacks} \cite{adv_rfml:Shi2019a}.
While a Spoofing Attack may provide benefits over a simpler \emph{Relay Attack} by considering channel and receiver effects in the adversarial transmission, no communication can occur using such an attack.
Therefore, the benefits of such an attack would be limited and the more prevalent threat is considering how signals can be manipulated without losing their underlying semantic meaning.

\paragraph{Defining Perceptible Perturbations in Wireless}
The main constraint in adversarial machine learning is generally provided, particularly in \ac{CV}, as a constraint on the perturbation power, a proxy for the notion of perceptibility of the perturbation (e.g. does this perturbation affect a human observer's judgement of the image, interpretation of the audio signal content, or reading of a sentence).
This notion is more easily defined in \ac{RFML} as the \ac{BER} at a receiver.
More specifically, because the receiver is blind to the perturbation being applied, \ac{BER} defines the perceptibility of the adversarial attack \cite{adv_rfml:Flowers2020a} (i.e. the more obvious the perturbation, the higher the \ac{BER}).
Therefore, recent work has created differentiable versions of receive chains, that allow for the \ac{BER} to be directly incorporated into the loss function of an adversarial attack \cite{adv_rfml:Flowers2019a, adv_rfml:Hameed2019a}.
Thus, even though attacks transferred from \ac{CV} may have lower utility in wireless communications due to their large impact on the wireless transmission, the attackers will continue to evolve specifically in the context of \ac{RFML}, leading to more sophisticated threats.
As such, defenses must be investigated that mitigate future threats to \ac{RFML} systems being deployed in high risk adversarial environments.


\subsubsection{Defense} \label{sss:adv:defense}


Defending against adversarial attacks can be roughly split into two categories, discussed further in the following subsections:
\begin{enumerate}
\item[i.] detecting an attack is occurring in order to take counter-measures, or 
\item[ii.] becoming robust to attacks by increasing the power of the perturbation required to cause a misclassification.
\end{enumerate}

While this section only focuses on the work that has been done specifically for \ac{RFML} in the context of adversarial evasion attacks, more general surveys on adversarial attacks and defenses are provided in \cite{adv:Chakraborty2018a, adv:Akhtar2019a}.
Further, a more general discussion of detecting whether an example is \emph{in distribution} is left to Section \ref{sec:trust}.

\paragraph{Detecting Attacks}
Detecting an attack can be thought of as a supplemental binary classification that determines \emph{is this example in or out of distribution?}
Two metrics are proposed in \cite{adv_rfml:Kokalj-Filipovic2019a} for detecting adversarial attacks on wireless communications.
The first uses the distribution of the \ac{PAPR} of the underlying signal along with the model's classification.
Since the \ac{PAPR} can be used as a signature for a given modulation, the work in \cite{adv_rfml:Kokalj-Filipovic2019a} tests whether the \ac{DNN} classification and \ac{PAPR} signature are in agreement on the classification; if not, then the example is assumed to be an adversarial example.
This test is specific to the \ac{RFML} task, \ac{AMC}, but agnostic of the model used.
The second test uses the distribution of the output probabilities of the \ac{DNN} to determine whether an example is in or out of distribution and is therefore agnostic of the task it is applied to.
However, performing statistical tests during inference can increase system complexity on an already \ac{SWaP} constrained \ac{RFML} system (discussed further in Section \ref{sec:deployment}) leading to increased classification latency and thus decreased bandwidths that can be sensed in real time.
Additionally, this additional check can be incorporated into the attack, once the attacker has become aware, and likely bypassed just as the original classification was \cite{adv:Carlini2017b}.
Therefore, pushing the defense methodology into the training stage of the \ac{DNN}, where the computational complexity can be handled off target and without a time constraint, is often beneficial.

\paragraph{Becoming Robust to Attacks}
The most widely used methodology for gaining robustness is adversarial training \cite{adv:Goodfellow2015a, adv:Kurakin2016a, adv:Madry2017a, adv:Shafahi2019a, adv:Tramer2017a}.
Adversarial training is simply the introduction of correctly labeled adversarial examples during training time by a known adversarial attack (such as \ac{FGSM}).
Another method that alters the training strategy of \acp{DNN} to confer robustness is to reduce the degrees of freedom of an attacker by lowering the input dimensionality.
The work in \cite{adv_rfml:Kokalj-Filipovic2019c} adopted both strategies for training and observed that it increased the model's robustness to \ac{FGSM} attacks.
The results of \cite{adv_rfml:Kokalj-Filipovic2019c} also showed that lowering the input dimensionality alone was sufficient to increase robustness to an \ac{FGSM} attack; however, that adversarial training decreased the number of training epochs needed to reach near perfect accuracy on legitimate examples.
This highlights that adversarial training is not only good for conferring robustness, but can also be used a data augmentation technique for \ac{RFML} (Section \ref{sec:data}).
However, no work (yet) has shown that an adversarially trained classifier would be robust to \emph{all} attack methodologies \cite{adv:Dong2018a}. 

\paragraph{Moving Towards Mathematically Rigorous Definitions of Robustness}
Given that many proposed defenses have been quickly proven to be inadequate, it is important to be overly cautious when evaluating a new methodology \cite{adv:Carlini2019a}.
In addition to evaluating defenses against a huge, and growing, list of adversarial attacks (using an open source library such as Cleverhans can help alleviate the development burden \cite{adv:Papernot2016b}) research has begun looking into provable robustness.
One such work is \cite{adv:Weng2018a} that uses the Lipschitz function of \acp{DNN} to provide a lower bound on the adversarial distance needed to cause a misclassification.
More generally, this concept is about whether the model can be trusted on real inputs, where the inputs are distorted by some perturbation, regardless of whether the perturbations are man-made or naturally occurring.
Therefore, this discussion is deferred to Section \ref{sec:trust}.

\subsection{Mitigation Through Standard Security Practices} \label{ss:ops}
Defending a \ac{RFML} system from attack does not have to only revolve around adversarial \ac{ML} based defenses.
By using standard cybersecurity best practices, an adversary can be forced to move down the Threat Model presented in Figure \ref{fig:threat_model} by having their knowledge of and access to the \ac{RFML} system limited.
Therefore, the attacks become much more difficult to successfully execute.

More specifically, most adversarial attacks and defenses are proposed and evaluated in a fully digital world (a digital attack in Figure \ref{fig:threat_model}); however, such attacks and defenses can transfer to the physical environment as well \cite{adv:Kurakin2016b}.
In the context of \ac{RFML}, this physical environment means that the perturbation is radiated from an external transmitter. 
Therefore, the transmission is impacted by channel effects, hardware impairments at both the transmitter and receiver, and \ac{DSP} pre-preprocessing techniques before reaching the \ac{DNN} for classification (a physical attack in Figure \ref{fig:threat_model}).
All of these can serve as an impediment for an attacker, forcing them to raise their adversarial perturbation power in order to overcome the effects \cite{adv_rfml:Flowers2020a, adv_rfml:Hameed2019a, adv_rfml:Kim2020a, adv_rfml:Usama2019a}.

Additionally, so-called \emph{white-box} attacks, which assume full knowledge of the target \ac{DNN}, are generally known to be more effective than \emph{black-box} attacks which assume close to zero knowledge about the target, regardless of modality. 
This is not meant to say that adversarial examples do not transfer between models, only that when transferring adversarial examples between models, a small penalty on the adversary's success is incurred.
However, limiting the amount of information an adversary can build up about the \ac{DNN} is a critical first step in defending against attack.


\subsection{Discussion and Future Work} \label{ss:adv_future}
The broader community's understanding of adversarial examples in terms of \emph{how} to create and defend against them, \emph{when} to inject them into a \ac{DNN} (training/inference), and \emph{why} they exist, is still rapidly evolving.
Much of this discussion can be applied generally across all data modalities; however, \ac{RFML} provides unique considerations that must also be studied separately:
\begin{enumerate}[i.]
    \item the physical channels between adversary and receiver are significantly different,
    \item the perceptability of the perturbation is machine defined, not human defined, and
    \item the actions taken based on the generated knowledge are application specific.
\end{enumerate}
Due to i. and ii., adversarial attacks from other modalities are of limited concern to deployed \ac{RFML} systems as the wireless channel forces the adversary to increase the perturbation power to a level that significantly interferes with the primary objective of the transmission: to communicate.
Ongoing work has shown that more sophisticated attacks will emerge to overcome these limitations by incorporating more expert knowledge (i.e. channel type \cite{adv_rfml:Kim2020b}, channel coding scheme\cite{adv_rfml:DelVecchio2020a}, device type\cite{adv_rfml:Restuccia2020b}) into the adversarial process.
Another key research direction going forward is determining what information is emitted from a \ac{RFML} device, such as acknowledgement messages or transmission decisions \cite{adv_rfml:Restuccia2020b, adv_rfml:Erpek2019a}, that can aid in online learning for increased attack effectiveness.
Finally, a major benefit of a \ac{RFML} controlled radio over a traditionally deterministic policy based approach is the ability of the radio to continually learn and adapt to its specific environment.
Thus, the recent work in determining how the training data of \ac{RFML} systems can be manipulated to cause a degradation in model performance \cite{adv_rfml:Davaslioglu2019a, adv_rfml:Shi2018a} will motivate the study of data cleaning methodologies for wireless.

\section{Trust and Assurance}\label{sec:trust}

For all \ac{RFML} applications (Section \ref{sec:applications}), there is a desire to translate laboratory \textit{performance} into a user-defined \textit{mission assurance}.
This is critical to not only assuring that the machine learned behaviors, which are difficult to reverse engineer, behave as expected when put to practice, but more importantly, understanding how the system will respond to unanticipated stimuli and/or recognizing that an input is outside the training set of its learned responses.  
Taking this need to an extreme, a significant amount of end-user confidence is required to give \ac{RFML} systems the authority to permit autonomous weapons release \cite{kania2020}.

In most current \ac{RFML} techniques evaluated, learned behaviors are a function of correlation rather than causation.
That is, algorithms are data-driven, and thereby assume that the training, validation, and test datasets used to develop said algorithm are drawn from the same distribution which will be seen once deployed.
A primary concern for early adopters of \ac{RFML} is how the algorithm will behave when this assumption is invalid, either due to the real-world considerations not present in the training data or adversarial attack as discussed in Sections \ref{sec:data}, \ref{sec:security}, and \ref{sec:deployment}. 

\begin{figure}[t]
	\centering
	\includegraphics[width=\linewidth]{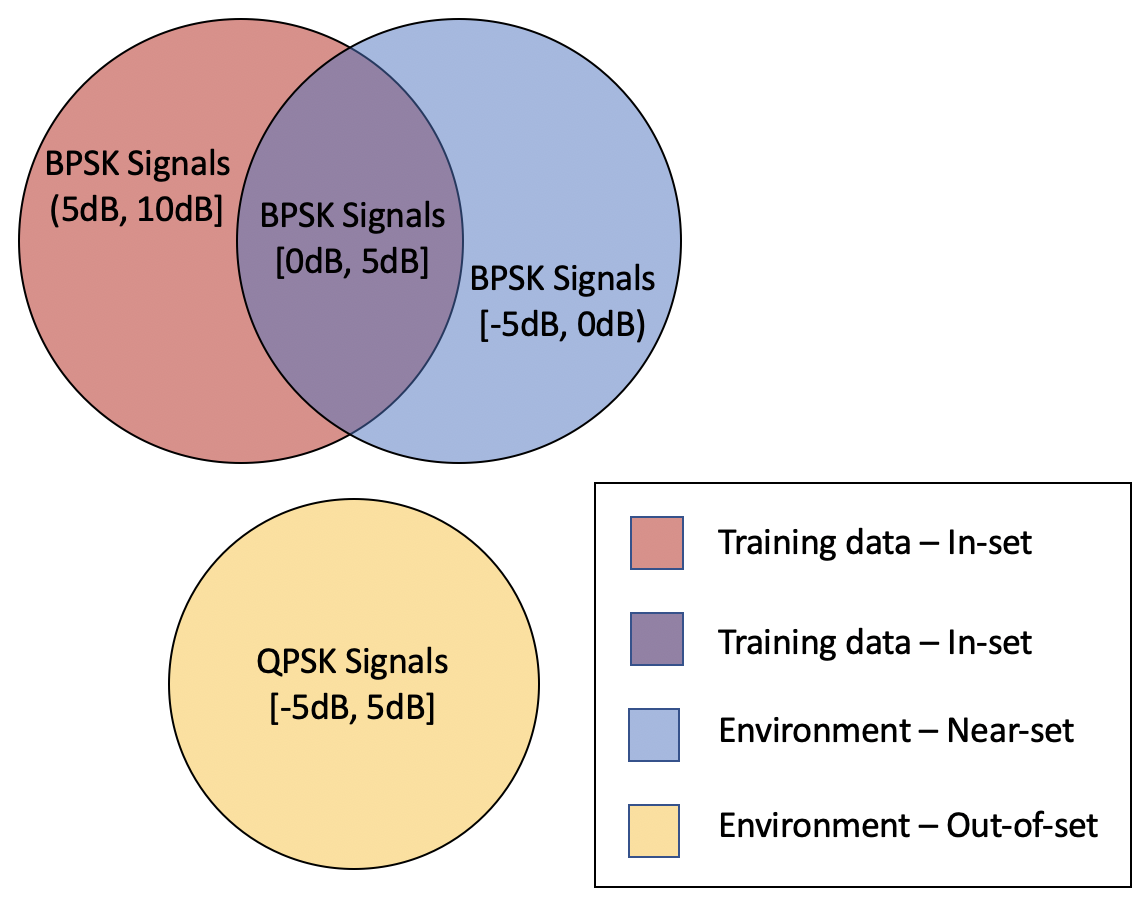}
	\caption{An example of in-set, near-set, and out-of-set data types for an \ac{RFML} algorithm trained on BPSK signals with SNRs between 0dB and 10dB.}
	\label{fig:venn}
\end{figure}

In other words, we can categorize inputs in one of three ways, illustrated in Figure \ref{fig:venn}:
\begin{itemize}
\item \emph{in-set} - Those that match the distribution of the training data. Using modulation classification as an example, an in-set input is a known modulation scheme under the same channel effects, SNR, transmitter/receiver imperfections, etc. that were seen during training.
\item \emph{near-set} - Those close to the distribution of the training dataset, but not included. In our modulation classification example, near-set inputs might be a trained modulation scheme, but different channel effects or SNR.
\item \emph{out-of-set} - Those completely outside of the distribution of the training data. Completing our example, an out-of-set example would be an untrained modulation scheme.
\end{itemize}
A data-driven \ac{RFML} system will behave as expected on \emph{in-set} inputs, but is unpredictable on \emph{near-set} input values, and necessarily incorrect in processing all \emph{out-of-set} input values.
This points to a critical need for approaches to both rigorously assure ``the safety and functional correctness" of \ac{RFML} systems throughout deployment \cite{darpa_aai} and explain or understand the behavior of RFML systems \cite{darpa_xai}.

These concerns are not unique to \ac{RFML} \cite{carvalho2019}, but have yet to be adequately addressed in \ac{RFML} literature.
Therefore, this section will focus on the very young body of work in testing, verification, and interpretation of general \ac{ML} systems which could be explored for use in the RF domain, and will discuss the pros and cons of the various approaches.
These works can broadly be categorized into three research areas,
\begin{itemize}
\item \emph{verification} methods which provide mathematical proof that a desired property holds for a trained model,
\item \emph{testing} methods which aim to exhaustively evaluate a trained model to identify flaws, and 
\item \emph{interpretation/explanation} techniques which include methods to describe and/or quantify a trained model's learned behaviors in a human-understandable format, such as decision/model explanations or uncertainty/reliability metrics.
\end{itemize} 

\subsection{Verification}
Beginning with the most rigorous approaches, current methods for verifying ML algorithms apply formal methods such as constraint solving \cite{katz2017_verification, bunel2018_verification, dvijotham2018_verification}, global optimization \cite{ruan2019_verification}, search-based methods \cite{wu2020_verification}, and over approximation \cite{gehr2018_verification} to provide provable guarantees about their behavior when provided with in-set, near-set, out-of-set, and even adversarial examples.
While verification methods provide deterministic or statistical guarantees of the robustness of previously trained models, they are also typically NP hard or extremely computationally expensive and time intensive.
As a result, none have been able to successfully scale to the state-of-the-art deep \acp{NN} used today.

Towards scalable \ac{ML} verification methods, a promising path forward is that of approximate or iterative/anytime methods which provide provable upper/lower bounds or monotonically converging bounds \cite{ruan2019_verification, dvijotham2018_verification, gehr2018_verification, wu2020_verification}.
However, future work is needed to improve these methods, in order to yield tighter and more useful bounds on the robustness of trained models.

\subsection{Testing}
Traditionally, \ac{ML} researchers and engineers rely on a held-out test set, which remains unseen throughout the training and model selection process, to provide an estimate of a trained model's performance \cite{kuhn2013_datasplit}.
This computationally efficient method provides a good estimate of how the model will perform on in-set data, but fails to identify how the model will perform on near-set or out-of-set data.

In an effort to strike a balance between computational efficiency and rigorousness, there is a growing body of work adapting and applying software testing and debugging techniques to more thoroughly test ML algorithms.
These approaches generate test cases using methods such as concolic testing \cite{sun2018a_testing, sun2018b_testing}, mutation testing \cite{ma2018a_testing}, differential analysis \cite{ma2018c_testing}, or even adversarial methods \cite{carlini2017}, which is typically guided by a user-selected coverage metric. 
Some of the most popular coverage metrics used have included neuron or layer coverage \cite{pei2017_testing, ma2018b_testing} and modified condition/decision coverage \cite{sun2018a_testing}.
The aim is to generate a set of test cases/inputs which provide sufficient coverage of the trained model, dictated by a user-selected threshold.

Though test case generation may provide more assurance than traditional \ac{ML} testing practices and are typically more computationally efficient than verification methods, there are a number of drawbacks which should be addressed.
First and foremost, like traditional software testing methods, \ac{ML} testing methods can only identify a lack of robustness, and cannot ensure robustness.
In the same vein, the effectiveness of the testing method is highly dependent upon the coverage metrics and thresholds used, both of which are chosen by the user. 
With some coverage metrics and thresholds, testing methods may be just as computationally expensive and time consuming in comparison to approximate verification methods.
Ultimately, while there is certainly value in more effective testing techniques, future work will likely need to focus on \ac{RFML} verification over \ac{RFML} testing, in order to effectively mitigate against adversarial attack and provide assured performance \cite{goodfellow2017_blog}.

\subsection{Interpretation/Explanation} 
Switching gears, the aim of interpretation/explanation methods is to address the challenge of \ac{HMI} by ``enabl[ing] users to understand how the data is processed and supports awareness of possible bias and systems malfunctions" \cite{mohseni2018}.
In other words, \ac{HMI} becomes more feasible if the model/decision is better understood by the end user.
Approaches to interpret or explain black-box ML models such as deep NNs and/or their decisions can broadly be categorized into two groups.

The first group of approaches provide \emph{intrinsic interpretability} by using inherently more interpretable models either from the offset or extracted from a black box model \cite{du2019}. Examples of such models include decision trees \cite{clark-amc, bastani2017decisiontree}, attention mechanisms \cite{bahdanau2014attention}, clustering algorithms, or linear/Bayesian classifiers \cite{molnar2019interpretable}.
While these methods are typically the most straightforward and provide the most useful model/decision explanations, inherently interpretable models are typically less expressive than black-box models such as deep \acp{NN}, and therefore do not provide the same level of performance. 

The second group of approaches provide \emph{post-hoc interpretability} by extracting decision/model explanations from black-box models or through model exploration \cite{du2019, mohseni2018}.
Post-hoc interpretability methods can be further broken down into \emph{local interpretability} methods and \emph{global interpretability} methods.
Local interpretability methods aim to provide an explanation for why and/or how a black box model made the decision it made for a given example input.
These instance-level explanations can be aggregated over a group of example inputs to draw larger conclusions about a model's knowledge.
Meanwhile global interpretability methods focus on increasing the transparency of black-box models by ``inspecting the structures and parameters" in an effort to understand the scope of the model's knowledge more directly \cite{du2019}. 

Local interpretability methods typically utilize some form of visualization to describe the network's response to the input such as heatmaps, which indicate which portions of the example input contributed most to the network's decision \cite{bach2015lrp, simonyan2013saliency}.
Popular and successful local interpretability methods in the image processing domain include backpropagation techniques such as layerwise relevance propagation, Taylor decomposition, and GradCAM \cite{bach2015lrp, springenberg2014guidedbackprop, selvaraju2017gradcam}, saliency mapping \cite{simonyan2013saliency}, and deconvolutional networks \cite{zeiler2014deconv}.
However, transitioning these methods to the \ac{RF} domain has proven challenging, as raw RF data is more difficult to visualize, especially in the intermediate layers of a deep \ac{NN}.
Therefore, a more promising local interpretability method for use in the RF domain is the use of uncertainty metrics to accurately quantify a model's confidence in any given decision, and could be used to identify unpredictability due to adversarial attack or operating environments \cite{gal2016_uncertainty, jha2019confidence}.

Global interpretability methods focus less on visualization techniques due to the large number of parameters in deep NN models, but have been explored through approaches such as activation maximization and partial dependence \cite{simonyan2013saliency, yosinski2015viscnn, hooker2004dependence, krause2016dependence2}.
More common is the use of metrics such as feature importance \cite{zien2009firm, vidovic2016firm2}, sensitivity \cite{saltelli2002sensitivity, olden2002sensitivity2}, and mutual information \cite{shwartz2017info}.

The primary challenge shared amongst both local and global interpretability methods is that there are no universal definitions for terms such as trust, interpretability, assurance, and explanation in the deep learning literature.
Furthermore, the concept of interpretability is highly dependent on the end user and their technical background \cite{lipton2018}.
For example, some argue that while global interpretability methods are useful to the deep learning expert who understands the inner-workings of a black-box model,  local interpretability methods are more tangible, intuitive, and provide more benefit to the end user.
Furthermore, trust, interpretability, assurance, and explanation are largely gauged qualitatively rather than quantitatively, and therefore are hard to compare and evaluate across approaches \cite{mohseni2018}.

Additional challenges to \ac{ML} interpretations include, but are not limited to \cite{carvalho2019}:
\begin{itemize}
\item How to accurately characterize and/or classify out-of-set examples. This is one area where uncertainty metrics would likely be more useful than visualization based explanation methods
\item Producing consistent explanations for similar inputs
\item Producing explanations without significant computational overhead
\item Deep \ac{NN} produce an overwhelming amount of highly complex and interdependent data that is difficult to visualize, describe, and/or explain in a helpful manner. The abstract nature of \ac{RF} data only exacerbates this challenge. 
\end{itemize}

\subsection{Discussion}
\emph{Trust/assurance} in \ac{RFML} systems will likely require some form of both verification method in conjunction with interpretation/explanation methods \cite{huang2018_trust}, in order to provide designer, administrator, operator, and end-user confidence in a model's decision-making capabilities both before and during deployment.
As discussed above, interpretation/explanation methods provide the user with an intuitive and/or quantifiable level of confidence in a model's decision, improving their understanding of and trust in the system.
While this understanding and trust is critical to \ac{HMI}, assured \ac{RFML} suitable for use in safety-critical systems, such as self-driving cars and military systems, will require the rigorous guarantees that verification provides.


\section{Deployment}\label{sec:deployment}
Early adoption of \ac{RFML} systems has already taken place in a variety of military systems \cite{Roy,rfmls,signaleye,NI-5G}, though a broader interest will exist for the rollout of commercial cellular \cite{Balevi,MaHuMa,Kafle}, the \ac{IoT} \cite{AlHajri,McGinthy,Chatterjee,Guerra}, and even satellite communications \cite{LiuMorton,LiuZhang,Moody}.
Given the breadth of applications identified for \ac{RF} machine learning techniques given in Section \ref{sec:applications}, there is a fundamental desire to transition the developed techniques to real systems.  
This section evaluates the practical constraints of size, weight, power, cost, and performance bounds to facilitate deployment.  

\subsection{Size, Weight, Power, and Cost (SWaP-C)}

Beginning with size and weight, it is notable that many \ac{DL} techniques employ significant computing infrastructures during their training phases, the scope of which makes them infeasible for training in the field; however, we are most concerned with the \ac{DL} algorithm's computational requirements when attempting to process incoming, unknown, data inputs once deployed post training \cite{DGX2}.  
One advantage of \ac{RFML} techniques, as compared to most \ac{CV} techniques, is that the sizes of the associate \acp{NN} and processing are drastically smaller -- in many cases 2-3 orders of magnitude smaller in terms of number of parameters and thus memory utilization.
Further, some \ac{RFML} implementations incorporate pre-calculated traditional signal processing techniques such as Fourier and wavelet transforms, cyclostationary feature estimators, and other expert features to serve as a more efficient feature that may be merged with machine learned behaviors \cite{Altiparmak,LiuMorton,Bowen}.    
Other research has focused on reduced precision implementations of machine learning structures as a method to gain computational efficiency \cite{Camus,FoxFaraone,Colbert}.
However, the use of online learning techniques in \ac{RF} scenarios requires real-time computational resources that are currently difficult to reduce to a mobile system \cite{Gwon,NguyenTran}, in addition to the challenges discussed in Section \ref{sec:data}.  

Given the highly effective miniaturization of digital electronics, a deployed system's weight is primarily driven by its power consumption and the associated batteries or heatsinks \cite{Hannan}.  
As such, real-time evaluation of signal detection \cite{Vinsen}, signal characterization \cite{signaleye}, and specific emitter identification \cite{McGinthy} have each been evaluated, with the latter two being implementable to a tactical/mobile use case, either through the assumption of vehicle power or a tightly regulated and small duty cycle.  
Driving system criteria for this power usage include the instantaneous bandwidth of the spectrum analyzed, the density of signals within the environment (affecting the number of calls to an \ac{RFML} function), implementation in software versus hardware, and the environment where the device is used.
The use of wake-up circuits for periodic/event-triggered execution of a \ac{RFML} function can be used to drastically reduce average power draw \cite{McGinthy}.
Additionally, the use of energy harvesting techniques, and RFML processing within those energy production envelopes, are of particular interest for battery-powered IoT and solar-powered satellites.  

Beyond SWaP, cost is typically the next important deployment consideration.  
To date, the primary cost of \ac{RFML} systems appears to be underlying datasets used for training, followed by the training process and hardware, and finally by the fieldsets to be deployed.  
As discussed previously in Section \ref{sec:data}, the quality of the data drives overall functionality, and often requires human-intensive labeling and/or pre-processing \cite{rfmls}.  
Training costs are primarily driven by the purchase and use of parallelized processors such as Graphics Processing Units, Tensor Processing Units, or other special purpose hardware.  
Both the data collection and training processes can be amortized over the number of fieldsets -- in current military deployments, the number of deployed units is typically small, making the relative cost per mobile fieldset high.  
By contrast, deployment for a widely used 5G/6G cellular application will be make the training costs, if broadly applicable, nearly negligible.

\subsection{Application Dependencies}

As described in Sections \ref{sec:applications} and \ref{sec:data}, the scale and scope of different applications can lead to vastly different hardware and SWaP requirements -- at one end of the spectrum a Raspberry Pi 0 for performing event-triggered packet-based \ac{SEI} for \ac{IoT} networks \cite{McGinthy} and on the other end a real-time 5 GHz instantaneous continuous spectrum monitoring system \cite{rfmls}.  
One specific application example, driven by environmental effects, is the potential deployment of \ac{ML} algorithms aboard small spacecraft, which are impacted by radiation-induced single event upsets \cite{arechiga-seu1,arechiga-seu2} -- without the addition of radiation shielding and/or extensive mitigation strategies, the performance of the \ac{ML} structures fail to achieve the necessary performance \cite{LiGuanpeng,altland-hack4,Reagen,Yan2019,Neggaz,Ozen} to be practically useful.
Broader dependencies include harnessing the more rapid decision making of \ac{RFML} - in many applications discussed in Section \ref{sec:applications}, the envisioned use case for \ac{RFML} is primarily as a decision aid.  Additional work will be required to make the raw observables fully actionable in live systems.

\subsection{Open Questions for Deployment}

In addition to those described previously in this section, we see the need for future research that addresses the following capabilities before \ac{RFML} algorithms can supplant existing techniques.

\begin{itemize}
    \item Online, incremental, and transfer learning techniques: current training processes generally assume up front training with a defined set of signal classes or RF environmental conditions.  Work is needed to add new signals or emitters without repeating the training process, as discussed previously in Section \ref{sec:data}.
    \item New processing capabilities: expanding upon work for reduced-precision implementations of neural nets, research is needed to evaluate more computationally efficient designs, such as pruning, bit-slice processor architectures for offline/intermittent calculation, potential insertion of \ac{RFML} math co-processors during execution stages, etc.
    \item Online learning: given the real-time nature of spectral observations, there is a need to improve the efficiency and timeline of online learning that absorbs new information into the trained behaviors.
    \item Distributed \ac{RFML}: most current work is focused on individual nodes, yet many \ac{RF} systems offer the potential of multiple apertures whose spectrum observations can be integrated to gain a larger system picture, as briefly mentioned in the context of transfer learning in Section \ref{sec:data}.  Example applications for expansion include multi-node geo-location and transfer learning using compressed samples or intermediate outputs of the RFML processing.
    \item Human interaction with \ac{RFML}: beyond individual decisions for signal detection or classification, additional work is needed in helping the end user understand the limits of the learned behaviors, how to shape and/or optimize use the system, and how to visualize and/or verify whether the machine should be trusted, a topic further explored in Section \ref{sec:trust}.
    \item Confidence: also discussed further in Section \ref{sec:trust} as well as in Section \ref{sec:security}, additional work is required to identify in real-time if prior training is truly representative of decisions requested as well as the quality of the resulting decision (i.e. consistency with laboratory-calculated performance) in order to provide assured performance, as well as to begin ruggedizing the decision chain against spoofing and other adversarial techniques 
\end{itemize}

\section{Conclusion}
As shown in Section \ref{sec:applications}, \ac{RFML} is a rapidly growing area of research, due to its demonstrated success in improving and automating spectrum sensing applications and supporting next-generation cognitive/intelligent radio applications. 
However, these works have primarily focused on conforming existing image or natural language processing solutions to an \ac{RF} application, each making their own assumptions about dataset availability, use cases, etc, and often ignoring key considerations, common to all \ac{RFML} solutions, that must be addressed in order to make deployable \ac{RFML} technologies realizable.

In this work, these common considerations, termed the ``RFML Ecosystem" were defined as the application itself (Section \ref{sec:applications}), dataset creation (Section \ref{sec:data}), security (Section \ref{sec:security}), trust and assurance (Section \ref{sec:trust}), and deployment (Section \ref{sec:deployment}).
For each element of the ecosystem, along with an overview of the topic, the primary research areas were identified with examples of existing works.
Additionally, discussion of dependencies between the elements of the RFML ecosystem provide a comprehensive and integrated summary of the domain-specific challenges to applying \ac{DL} to \ac{RF}.
In whole, this work aims to be a holistic guide for \ac{RFML} developers looking to develop realizable and deployable solutions for real-world applications and to promote the advancement of \ac{ML} architectures and algorithms purpose-built for the \ac{RF} domain.

\bibliographystyle{IEEEtran}
\bibliography{references/main.bib,references/adv.bib,references/adv_rfml.bib}
\balance
\end{document}